\documentclass[twocolumn,aps,prl,groupedaddress,showpacs,superscriptaddress,floatfix]{revtex4-2} 

\usepackage{amsmath}
\usepackage{amsfonts}
\usepackage{upgreek}
\usepackage{amssymb}
\usepackage{braket}
\usepackage{graphicx,bm,units,yfonts}
\usepackage[table]{xcolor}
\usepackage{color}
\usepackage[colorlinks,citecolor=blue,urlcolor=red,bookmarks=false,hypertexnames=true]{hyperref}

\usepackage[utf8]{inputenc}

\usepackage{amstext}

\makeatletter
\let\openbox\@undefined
\makeatother
\usepackage{amsthm}

\makeatletter

\theoremstyle{plain}

\theoremstyle{remark}

\renewcommand{\vec}[1]{\mathbf{#1}}

\makeatother

\providecommand{\remarkname}{Remark}
\providecommand{\theoremname}{Theorem}

\begin{document}

\title{Classifying topology in photonic heterostructures with gapless environments}

\author{Kahlil Y. Dixon}
\email{kydixon@sandia.gov}
\affiliation{Center for Integrated Nanotechnologies, Sandia National Laboratories, Albuquerque, New Mexico 87185, USA}

\author{Terry A.\ Loring}
\affiliation{Department of Mathematics and Statistics, University of New Mexico, Albuquerque, New Mexico 87131, USA}

\author{Alexander Cerjan}
\email{awcerja@sandia.gov}
\affiliation{Center for Integrated Nanotechnologies, Sandia National Laboratories, Albuquerque, New Mexico 87185, USA}

\date{\today}

\begin{abstract}
Photonic topological insulators exhibit bulk-boundary correspondence, which requires that boundary-localized states appear at the interface formed between topologically distinct insulating materials. However, many topological photonic devices share a boundary with free space, which raises a subtle but critical problem as free space is gapless for photons above the light-line. Here, we use a local theory of topological materials to resolve bulk-boundary correspondence in heterostructures containing gapless materials and in radiative environments. In particular, we construct the heterostructure's spectral localizer, a composite operator based on the system's real-space description that provides a local marker for the system's topology and a corresponding local measure of its topological protection; both quantities are independent of the material's bulk band gap (or lack thereof). Moreover, we show that approximating radiative outcoupling as material absorption overestimates a heterostructure's topological protection. As the spectral localizer is applicable to systems in any physical dimension and in any discrete symmetry class, our results show how to calculate topological invariants, quantify topological protection, and locate topological boundary-localized resonances in topological materials that interface with gapless media in general.
\end{abstract}

\maketitle

Recent advances in topological photonics \cite{lu_topological_2014,khanikaev_two-dimensional_2017,ozawa_topological_2019} have led to the development of novel technologies including topological lasers~\cite{bahari_nonreciprocal_2017,st-jean_lasing_2017,bandres_topological_2018,zeng_electrically_2020,yang_spin-momentum-locked_2020,shao_high-performance_2020,bahari_photonic_2021,dikopoltsev_topological_2021,yang_topological-cavity_2022} and devices that create and route quantum light~\cite{rechtsman_topological_2016,barik_topological_2018,mittal_topological_2018,barik_chiral_2020,parappurath_direct_2020,arora_direct_2021,dai_topologically_2022,hauff_chiral_2022}. However, the utility of many of these devices is predicated on the presence of, and potential coupling to, scattering channels in the surrounding environment that are degenerate with the boundary-localized topological states that underpin these devices' functionality. Thus, even though these devices can feature photonic crystals or other lattices with complete topological band gaps in their interior, their boundary-localized states are generally resonances, not bound modes, which radiate into the surrounding environment as free space is gapless above the light line.

\begin{figure}
    \centering
    \includegraphics{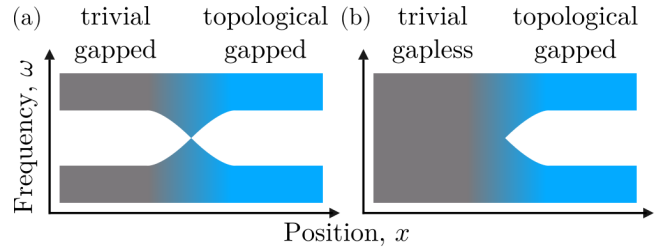}
    \caption{(a) Schematic of the local density of states as the probed position is varied across the interface of a heterostructure formed by a trivial insulator and topological insulator with a common bulk band gap. (b) A similar schematic, except in which the trivial material is gapless.}
    \label{fig:1}
\end{figure}

Unfortunately, the fact that the typical environment for topological photonic structures is gapless, rather than gapped (i.e., insulating), presents a fundamental challenge to our understanding of these devices. Heuristically, topological boundary-localized modes form at the interface between two gapped materials with different bulk invariants as a resolution to the need for band continuity across the heterostructure's interface; the band gap must close in the vicinity of the interface so that the different bulk invariants can be reconciled, yielding interface-localized states \cite{lu_topological_2014}, see Fig.\ \ref{fig:1}a. Indeed, traditional approaches to material topology have been highly successful at predicting the interface phenomena in heterostructures featuring topologically non-trivial insulators~\cite{hasan_colloquium:_2010, qi_topological_2011,bansil_colloquium_2016,chiu_classification_2016,ozawa_topological_2019} and semimetals~\cite{wang_dirac_2012, liu_discovery_2014, wang_three-dimensional_2013,weng_weyl_2015, lv_observation_2015, lu_experimental_2015, xiao_synthetic_2015, li_weyl_2018, xie_experimental_2019, yang_topological_2019_triply,  burkov2011topological, xu2015two, fu2019dirac}. But, if at least one of the materials in a heterostructure is gapless, this explanation fails, as the band gap must close in the vicinity of the interface regardless (so as to satisfy bulk band continuity between the two materials); any need to reconcile different bulk material topologies could occur as part of this standard band closing process without resulting in topological interface-localized states or resonances, see Fig.\ \ref{fig:1}b. Note, in this context ``gapless'' specifically refers to a $d$-dimensional material with $(d-1)$-dimensional isofrequency contours over a given range of wavelengths (i.e., those wavelengths in the other material's bulk band gap).
Thus, the plethora of photonic experiments that have observed topological boundary-localized resonances in devices that abut and radiate to free space suggests that material topology must be definable in heterostructures containing a gapless material, even if the lack of a global bulk band gap prohibits the use of traditional theories of physical topology.

Here, we identify topological boundary-localized resonances and quantify their protection in gapless heterostructures with radiative environments using a theory of topological materials based on their real-space description. To do so, we construct the heterostructure's spectral localizer, a composite operator that combines a system's Hamiltonian and position operators with a Clifford representation, and which provides local topological markers and a spatially resolved measure of protection even for non-Hermitian systems. We demonstrate this topological classification approach on a 2D photonic Chern crystal embedded in free space with radiative boundary conditions. Using this model, we also show that radiative losses and material absorption have qualitatively different consequences for a system's topological protection, and approximating radiative outcoupling as absorption will substantially overestimate the protection of the boundary-localized resonances. Finally, we provide an example of how topological robustness against system disorder manifests in this real-space classification approach. Our results prove that bulk-boundary correspondence is still required in gapless heterostructures, providing a rigorous framework for understanding many types of topological photonic devices. 

\begin{figure}
    \centering
    \includegraphics{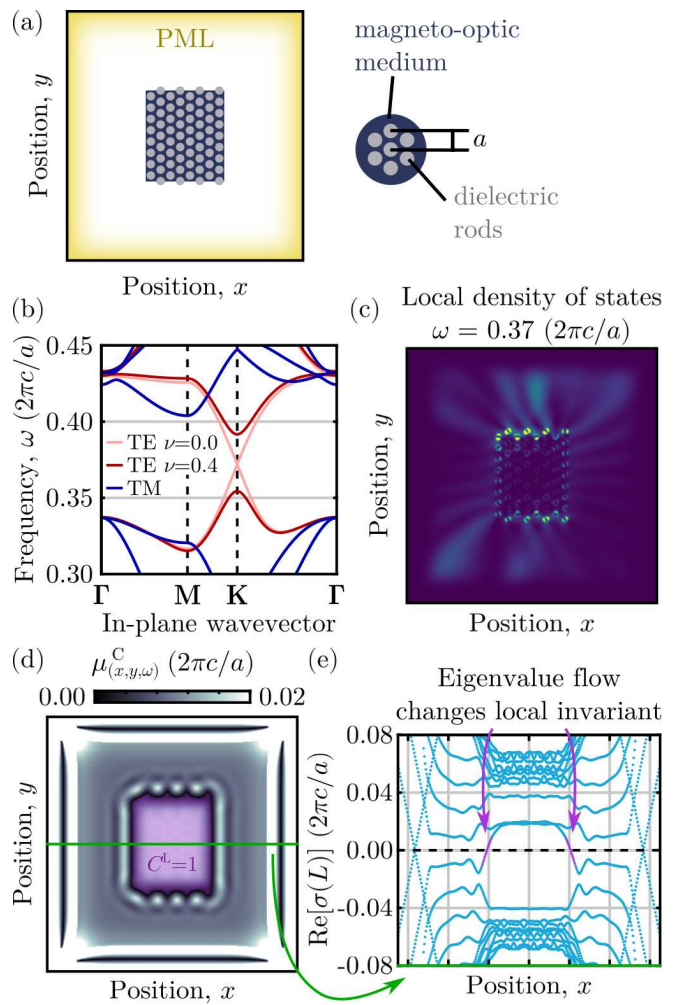}
    \caption{(a) Schematic of a 2D photonic Chern insulator embedded in free space $\varepsilon_{\textrm{fs}} = 1$ with radiative boundary conditions. The topological photonic insulator is comprised of dielectric rods $\varepsilon_{\textrm{rod}} = 14$ with spacing $a$ in a magneto-optic background $\varepsilon_{\textrm{mo}} = \big(\begin{smallmatrix}
     1 & -i\nu \\
     i \nu & 1
    \end{smallmatrix}\big)$. (b) Bulk band structure for the photonic Chern insulator for the TE modes with $\nu = 0$ (light red) and $\nu = 0.4$ (dark red), and the transverse magnetic (TM) modes that are independent of $\nu$. (c) Local density of states for the finite system at $\omega = 0.37 (2\pi c/a)$. (d) Local gap $\mu^{\textrm{C}}$ with overlaid local index ($C^{\textrm{L}} = 1$ is shown as magenta) for the finite system at $\omega = 0.37 (2\pi c/a)$. Both (c) and (d) are shown on the same spatial scale as (a). (e) Spectral flow of the real parts of the 20 eigenvalues of $L$ closest to $0$ for $y$ fixed to the center of the finite system (shown as the green line in (d)), and at $\omega = 0.37 (2\pi c/a)$. (d) and (e) are calculated using $\kappa = 0.04 (2 \pi c/a^2)$.}
    \label{fig:2}
\end{figure}

We begin by considering a prototypical topological photonic system consisting of a photonic Chern insulator embedded in free space. In particular, we use a finite portion of the 2D magneto-optic photonic crystal proposed by Haldane and Raghu \cite{haldane_possible_2008,raghu_analogs_2008} surrounded on all sides by vacuum, with the radiative boundary condition implemented using stretched-coordinate perfectly matched layers (PML) \cite{taflove_advances_2013}, see Fig.\ \ref{fig:2}a. When an external magnetic field is applied, a topologically non-trivial band gap opens in the photonic crystal's transverse electric (TE) sector that supports chiral edge modes within this gap, see Fig.\ \ref{fig:2}b. As the photonic crystal in our model system is finite, all of its states, including its chiral edge modes, are resonances as they decay due to radiative outcoupling. These chiral edge resonances can be seen in the system's local density of states (LDOS) within the bulk band gap of the photonic Chern insulator, Fig.\ \ref{fig:2}c. Altogether, this model system preserves all of the salient features of many topological photonic systems that have been previously experimentally observed \cite{rechtsman_photonic_2013,hafezi_imaging_2013}, but whose topological protection cannot be quantified using topological band theory because the materials that form the heterostructure lack a common bulk band gap.

Instead, to prove that the gapless heterostructure in Fig.\ \ref{fig:2}a must possess protected boundary-localized resonances due to the non-trivial topology of the central lattice, we employ the spectral localizer \cite{loring_k-theory_2015,loring_finite_2017,loring_spectral_2020}. For a $d$-dimensional system, the spectral localizer is a composite operator that combines the system's Hamiltonian $H$ and position operators $X_1, X_2,..., X_d$ using a non-trivial Clifford representation, and yields both a local topological marker and local measure of protection. For the non-Hermitian 2D system that we consider here, we can use the Pauli matrices as the Clifford representation (as they generate a representation of $\mathcal{C}\ell_3(\mathbb{C})$) to write the spectral localizer as \cite{cerjan2023spectral}
\begin{align} 
    &{L}_{(x,y,\omega)} (X,Y,H) = \label{eq:L} \\ 
        &\left( \begin{array}{cc}
        H - \omega I & \kappa(X-xI) - i\kappa(Y-yI) \\
        \kappa(X-xI) + i\kappa(Y-yI) & -(H - \omega I)^\dagger
        \end{array} \right). \notag
\end{align} 
Here, $x,y,\omega$ are the choices of position and frequency where the spectral localizer is evaluated, $X$ and $Y$ are the 2D position operators, $I$ is the identity matrix, and $\kappa$ is a positive scaling coefficient with units of frequency times inverse distance.

Intuitively, the spectral localizer can be viewed as a composition of the eigenvalue equations (such as $(H-\omega I)|\psi \rangle = 0$) of the (generally) non-commuting operators $X,Y,H$ using the Pauli matrices. Despite the lack of a joint spectrum for $X,Y,H$, the spectral localizer can be used to determine whether a given choice of $x,y,\omega$ yields an approximate joint eigenvector of $X,Y,H$, i.e.\ is there some vector $|\phi \rangle$ for which $H|\phi \rangle \approx \omega |\phi \rangle$, $X|\phi \rangle \approx x |\phi \rangle$, and $Y|\phi \rangle \approx y |\phi \rangle$ \cite{cerjan_quadratic_2022}. A measure of how good these approximations are is given by
\begin{equation}
    \mu_{(x,y,\omega)}^\textrm{C}(X,Y,H) = \textrm{min}( \vert \textrm{Re}[\sigma({L}_{(x,y,\omega)} (X,Y,H))]\vert ), \label{eq:mu}
\end{equation}
i.e., minimum distance over all of the eigenvalues of $L$ from the imaginary axis, where $\sigma(L)$ is the spectrum of $L$. Here, smaller values of $\mu_{(x,y,\omega)}^\textrm{C}$ indicate that $x,y,\omega$ are closer to yielding a joint eigenvector of $X,Y,H$, and those $x,y,\omega$ where $\mu_{(x,y,\omega)}^\textrm{C} \le \epsilon$ define the system's Clifford $\epsilon$-pseudospectrum \cite{loring_k-theory_2015,cerjan_quadratic_2022} (the superscript $\textrm{C}$ denotes Clifford). Note that even if $\mu_{(x,y,\omega)}^\textrm{C}(X,Y,H) = 0$, these approximations do not become exact.

A physical picture of the spectral localizer's connection to material topology can be built from the behavior of atomic limits. In an atomic limit, $[H^{(\textrm{AL})},X_j^{(\textrm{AL})}] = 0$, which stems from the system's Wannier functions being localized to a single lattice site \cite{kitaev_periodic_2009}. This commutation relation, coupled with the fact that position operators commute $[X_j,X_l] = 0$, requires the eigenvalues of ${L}_{(x,y,\omega)}$ to be equally partitioned between having positive and negative real parts for any choice of $x,y,\omega$ (i.e., for atomic limits $\textrm{sig}({L}_{(x,y,\omega)}) = 0$, where $\textrm{sig}$ denotes a matrix's signature, its number of eigenvalues with positive real parts minus its number with negative real parts) \cite{choi_almost_1988}. However, just as 0D systems can be topologically classified based on the number of eigenvalues they possess above and below a specified band gap \cite{kitaev_anyons_2006}, 2D systems can be locally classified based on the partitioning of the spectrum of ${L}_{(x,y,\omega)}$, assuming that $\mu_{(x,y,\omega)}^{\textrm{C}} > 0$ \cite{loring_k-theory_2015}.

Thus, if a generic system with $[H,X_j] \ne 0$ has $\textrm{sig}({L}_{(x,y,\omega)}) = 0$, then it is continuable to an atomic limit via a path of invertible matrices for that choice of $x,y,\omega$, i.e., the system is locally topologically trivial. Conversely, if $\textrm{sig}({L}_{(x,y,\omega)}) \ne 0$, there is an obstruction to finding such a path, and the system is topologically non-trivial at that $x,y,\omega$. As this classification approach is not restricting the matrix continuation path to preserve any system symmetries, the signature of $L$ defines a local Chern marker,
\begin{equation}
    C_{(x,y,\omega)}^{\textrm{L}}(X,Y,H) = \tfrac{1}{2}\textrm{sig}[L_{(x,y,\omega)}(X,Y,H)] \in \mathbb{Z}. \label{eq:CL}
\end{equation}
Moreover, as the partitioning of the spectrum of $L_{(x,y,\omega)}$ cannot change without $\mu_{(x,y,\omega)}^{\textrm{C}} = 0$, $\mu_{(x,y,\omega)}^\textrm{C}$ is a measure of the topological protection in a system and can be thought of as a ``local band gap.''

Altogether, the spectral localizer can be understood as a method for performing dimensional reduction consistent with Bott periodicity \cite{loring_k-theory_2015}. After dimensional reduction, the local invariants for all ten discrete symmetry classes become essentially one of the three invariants introduced by Kitaev \cite{kitaev_anyons_2006}, i.e., matrix signatures for $\mathbb{Z}$ invariants, or signs of determinants or signs of Pfaffians for $\mathbb{Z}_2$ invariants.

To apply the spectral localizer to the photonic crystal heterostructure considered in Fig.\ \ref{fig:2}a, we first reformulate Maxwell's equations into a Hamiltonian, with
\begin{gather}
    H(\vec{x}) = M^{-1/2}(\vec{x}) W M^{-1/2}(\vec{x}), \label{eq:Ham} \\
    W = \left( \begin{array}{cc}
   0 & -i \nabla \times \\
    i \nabla \times & 0
    \end{array} \right), \;\; \textrm{and} \;\; 
    M(\vec{x}) = \left( \begin{array}{cc}
    \overline{\overline{\mu}}(\mathbf{x}) & 0 \\
    0 & \overline{\overline{\varepsilon}}(\mathbf{x}) 
    \end{array} \right). \notag
\end{gather}
In doing so, we are assuming that the frequency dependence of the permittivity $\overline{\overline{\varepsilon}}$ and permeability $\overline{\overline{\mu}}$ tensors can be neglected over the frequency range of interest, and that both are semi-positive definite \cite{cerjan_operator_Maxwell_2022}. To use this Hamiltonian in Eq.\ (\ref{eq:L}), it must be discretized so that it becomes a bounded, finite matrix. Here, we use a standard 2D Yee grid \cite{yee_numerical_1966}. The discretization scheme also defines the position operators $X,Y$, which in the basis of Eq.\ (\ref{eq:Ham}) are diagonal matrices whose elements $[X]_{jj}$ and $[Y]_{jj}$ correspond to the spatial coordinates of the $j$th vertex in the discretization. Note that the stretched-coordinate PML makes $W$ non-Hermitian.

Overall, the spectral localizer numerical approach is similar to frequency-domain methods for solving Maxwell's equations because only a single frequency is considered within a given simulation. However, as Eq.\ (\ref{eq:L}) also requires specifying $x,y$ for each simulation and $L_{(x,y,\omega)}$ is connected to the approximate joint eigenvectors of $X,Y,H$, the spectral localizer approach is better classified as a ``pseudospectral-domain'' method. Thus, our implementation of Eq.\ (\ref{eq:L}) is a finite-difference pseudospectral-domain (FDPD) method and is publicly available \cite{GitHub,SuppInfo}.

Applying the spectral localizer to the topological photonic system considered in Fig.\ \ref{fig:2}a shows that for frequencies within the topological band gap of the photonic Chern crystal, the local gap $\mu^{\textrm{C}}$ closes around the boundary of the crystal, inside which the local Chern number becomes non-trivial $C_{(x,y,\omega)}^{\textrm{L}} = 1$, see Fig.\ \ref{fig:2}d. Moreover, monitoring the spectrum of $L$ near zero as one of the coordinates is varied across the system (fixing the other coordinate and $\omega$) directly shows the lone eigenvalue of $L$ responsible for the change in the system's local Chern marker, see Fig.\ \ref{fig:2}e. As locations where $\mu_{(x,y,\omega)}^{\textrm{C}} \approx 0$ indicate the presence of an approximate eigenstate of $H$ with eigenvalue near $\omega$ that is simultaneously approximately localized near $x,y$, the local gap closing around the topological photonic crystal can be understood as the manifestation of bulk-boundary correspondence in the spectral localizer framework. Thus, the spectral localizer proves that the boundary-localized resonances observed in topological photonic systems embedded in free space stem from the non-trivial topological material, and the fact that $\mu_{(x,y,\omega)}^{\textrm{C}} > 0$ in free space away from the interface (despite free space's gaplessness) is a measure of topological protection for this state. 

The spectral localizer's dependence on the choice of $\kappa$ in Eq.\ (\ref{eq:L}) can initially appear problematic. Indeed, for $\kappa = 0$, $L$ is block-diagonal, and its spectrum is always evenly partitioned such that $C^{\textrm{L}} = 0$. Conversely, for $\kappa \gg 1$, $L$ simply reveals the (exact) joint spectrum of $X$ and $Y$. However, in between these two limits there is a broad range of $\kappa$ over which a material's topological properties can be correctly predicted and remain effectively constant. In insulators, such a range always exists \cite{loring_finite_2017}. Moreover, in practice we find for our model system that $\kappa$ can be varied over more than two orders of magnitude while $C_{(x,y,\omega)}^{\textrm{L}}$ remains unaffected and $\mu_{(x,y,\omega)}^{\textrm{C}}$ only varies over a factor of two, see Supplemental Materials \cite{SuppInfo}.

\begin{figure}
    \centering
    \includegraphics{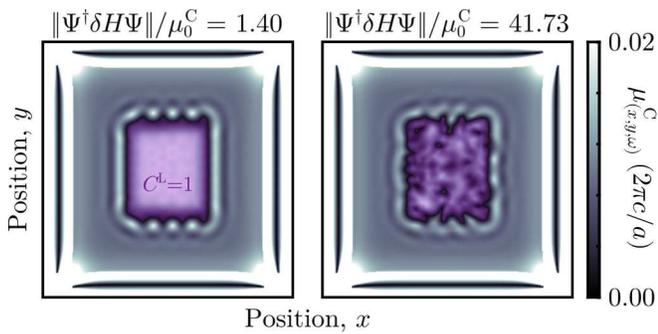}
    \caption{Local gap and overlaid topological marker ($C^\textrm{L} = 1$ shown in magenta) for the system shown in Fig.\ \ref{fig:2}a with added disorder with strengths $\Vert \Psi^\dagger \delta H \Psi \Vert/ \mu_0^{\textrm{C}} = 1.40$ (left) and $\Vert \Psi^\dagger \delta H \Psi \Vert/ \mu_0^{\textrm{C}} = 41.73$ (right) relative to the local gap at the center of the ordered system $\mu_0^{\textrm{C}} = 0.0185 (2\pi c/a^2)$. Disorder has been added to the high dielectric rod positions and dimensions and the disorder strength is calculated using the $m=370$ eigenvectors of $H$ closest to $\omega$, see Supplemental Materials \cite{SuppInfo}. Both figures are shown using the same spatial scale as Fig.\ \ref{fig:2}d, with $\omega = 0.37 (2 \pi c / a)$ and $\kappa = 0.04 (2\pi c/a^2)$.}
    \label{fig:3}
\end{figure}

Having proven that the chiral edge resonances seen in Fig.\ \ref{fig:2} are of topological origin, we now demonstrate their topological protection. In general, a system's topology at $x,y,\omega$ cannot change without $\mu_{(x,y,\omega)}^{\textrm{C}} \rightarrow 0$, as the local gap must close for one (or more) of the spectral localizer's eigenvalues to cross the imaginary axis. For Hermitian systems (which are Lipschitz continuous), one can prove that for a system perturbation $\delta H$ to close the local gap $\mu_{(x,y,\omega)}^{\textrm{C}}(X,Y,H+\delta H) = 0$, this perturbation must be at least as strong as the local gap is wide $\Vert \delta H \Vert \ge \mu_{(x,y,\omega)}^{\textrm{C}}(X,Y,H)$ \cite{loring_k-theory_2015}. For non-Hermitian line-gapped systems such as the one we consider here, this same criteria approximately holds \cite{cerjan2023spectral}. However, this known limit is not useful for evaluating the topological protection of photonic systems. The problem is that Maxwell's equations (prior to discretization) represent an unbounded operator, for which the $\ell^2$ norm is undefined. Thus, after discretization, even relatively modest perturbations will generally still yield substantial changes in the eigen-frequency of at least one high frequency state, yielding a large $\Vert \delta H \Vert$. Intuitively, the challenge is that $\mu_{(x,y,\omega)}^{\textrm{C}}$ is a local measure of protection in both position and frequency, yet $\Vert \delta H \Vert$ is a global measure of the perturbation.

Here, we conjecture that the correct measure of a perturbation's local strength is to project it into a subspace near $x,y,\omega$. For our model system, let $\Psi$ be an $n$-by-$m$ matrix whose $m$ columns are the eigenvectors of $H$ (which is $n$-by-$n$) with eigenvalues that are closest to $\omega$ where $\mu_{(x,y,\omega)}^{\textrm{C}}$ is calculated. Then, the local marker at $x,y,\omega$ cannot change so long as $\Vert \Psi^\dagger \delta H \Psi \Vert \lesssim \mu_{(x,y,\omega)}^{\textrm{C}}(X,Y,H)$. In Fig.\ \ref{fig:3} we provide numerical evidence for this conjecture by adding disorder to the positions and shapes of the photonic Chern insulator's high-dielectric rods in Fig.\ \ref{fig:2}a and calculating the disordered system's local topology. As can be seen, the predicted lower bound of topological protection holds; in fact, the local gap substantially underestimates the system's topological protection against this form of disorder. To instead demonstrate that our predicted bound can be saturated, one can remove the external magnetic field, $\nu = 0$, which both makes the system topologically trivial and corresponds to $\Vert \Psi^\dagger \delta H \Psi \Vert/ \mu_0^{\textrm{C}} = 1.81$.

\begin{figure}
    \centering
    \includegraphics{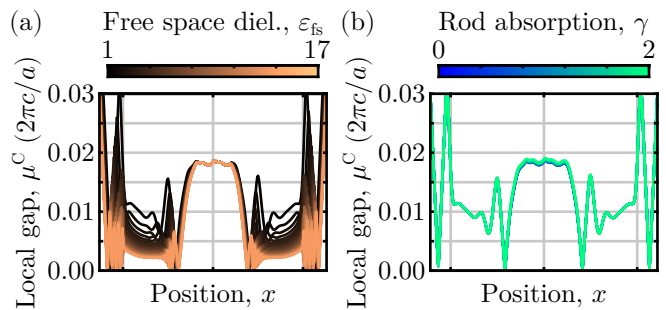}
    \caption{(a) Local gap for the system shown in Fig.\ \ref{fig:2}a for increasing values of the free space environment's dielectric $\varepsilon_{\textrm{fs}}$. (b) Similar to (a), except for increasing values of material absorption in the high-dielectric rods of the photonic Chern insulator, $\varepsilon_{\textrm{rod}} = 14 + i \gamma$, and with $\varepsilon_{\textrm{fs}} = 1$. Both figures are plotted along the green path in Fig.\ \ref{fig:2}d, with $\omega = 0.37 (2 \pi c / a)$ and $\kappa = 0.04 (2\pi c/a^2)$.}
    \label{fig:4}
\end{figure}

Beyond predicting a system's topological protection against crystal imperfections, the spectral localizer can also be used to approximate a system's robustness to surface roughness. In particular, while crystal imperfections serve to decrease the system's bulk band gap, and thus the effects of such perturbations can be captured using topological band theory, the dominant effect of surface roughness is to increase a system's radiative outcoupling. Thus, surface roughness cannot be considered without having a measure of topological protection for heterostructures lacking a global band gap. Here, we artificially increase our model system's radiative outcoupling by increasing the dielectric constant of the surrounding free space environment $\varepsilon_{\textrm{fs}}$. As can be seen in Fig.\ \ref{fig:4}a, even for values of $\varepsilon_{\textrm{fs}}$ greater than any material in the photonic Chern insulator, the spectral localizer is still able to predict the topology of the crystal, as well as the decreasing robustness of the chiral edge state. Given the connection between $\mu^{\textrm{C}}$ and the approximate joint spectrum of the system's operators, the decreasing local gap outside of the system for increasing $\varepsilon_{\textrm{fs}}$ is a manifestation of the increasing support (i.e., decreasing localization) of the chiral edge resonances outside of the photonic crystal. In contrast, if one instead approximates radiative outcoupling as material absorption, the topological protection of the chiral edge resonances is overestimated, see Fig.\ \ref{fig:4}b, as this approximation does not properly capture the salient physics that the system's chiral edge resonances are leaking out of the system's boundaries.

In conclusion, we have proven that gapless topological heterostructures still exhibit bulk-boundary correspondence despite the absence of a global band gap and have shown how to determine the protection of the resulting interface-localized resonances even in radiative environments. As such, we have resolved a subtle outstanding challenge in our understanding of topological photonics, where previous approaches to classifying devices embedded in air and operating above the light line required approximating free space to be trivially gapped. Through the spectral localizer, we show that treating radiative outcoupling as material absorption overestimates a system's topological protection. As the study of topological photonics turns towards developing devices for specific applications, the spectral localizer's ability to accurately predict topological robustness in radiative environments may enable new photonic device designs that are better protected against radiative outcoupling. Although we have presented this classification approach in a photonic systems, it is broadly applicable to topological materials in general, and in the Supplemental Materials we provide examples of using the spectral localizer to classify topology in gapless heterostructures formed from tight-binding models \cite{SuppInfo}.

\begin{acknowledgments}
A.C., T.L., and K.Y.D.\ acknowledge support from the Laboratory Directed Research and Development program at Sandia National Laboratories. T.L.\ acknowledges support from the National Science Foundation, grant DMS-2110398. K.Y.D.\ acknowledges support from the U.S.\ Department of Energy, Office of Basic Energy Sciences, Division of Materials Sciences and Engineering (BES 20-017574). This work was performed, in part, at the Center for Integrated Nanotechnologies, an Office of Science User Facility operated for the U.S. Department of Energy (DOE) Office of Science. Sandia National Laboratories is a multimission laboratory managed and operated by National Technology \& Engineering Solutions of Sandia, LLC, a wholly owned subsidiary of Honeywell International, Inc., for the U.S.\ DOE's National Nuclear Security Administration under contract DE-NA-0003525. The views expressed in the article do not necessarily represent the views of the U.S.\ DOE or the United States Government.
\end{acknowledgments}


%

\end{document}


\title{Supplemental Material for Classifying topology in photonic heterostructures with gapless environments}

\author{Kahlil Y. Dixon}
\email{kydixon@sandia.gov}
\affiliation{Center for Integrated Nanotechnologies, Sandia National Laboratories, Albuquerque, New Mexico 87185, USA}

\author{Terry A.\ Loring}
\affiliation{Department of Mathematics and Statistics, University of New Mexico, Albuquerque, New Mexico 87131, USA}

\author{Alexander Cerjan}
\email{awcerja@sandia.gov}
\affiliation{Center for Integrated Nanotechnologies, Sandia National Laboratories, Albuquerque, New Mexico 87185, USA}


\maketitle

\section{The role of the scaling coefficient $\kappa$}

In the spectral localizer, Eq.\ (1) of the main text, $\kappa$ serves two roles: it ensures that the whole matrix has consistent units, and it serves to tune the spectral localizer. This tuning is necessary because the two simple limits of $\kappa$ are not useful. When $\kappa = 0$, the spectrum of the spectral localizer is always equally partitioned about the imaginary axis as in this limit $L$ is block diagonal, with blocks $(H-\omega I)$ and $-(H-\omega I)^\dagger$. In the opposite limit, when $\kappa \rightarrow \infty$, the spectrum of $L$ simply identifies the joint spectrum of $X$ and $Y$, which commute. This spectrum of $L$ is also always symmetric about the imaginary axis, for similar reasons as before. From this perspective, it is somewhat remarkable that there is any regime of validity where the spectral localizer approach works at all, given that the two easily computable limits do not contain any new information about the system in question.

\begin{figure}[h]
    \centering
    \includegraphics{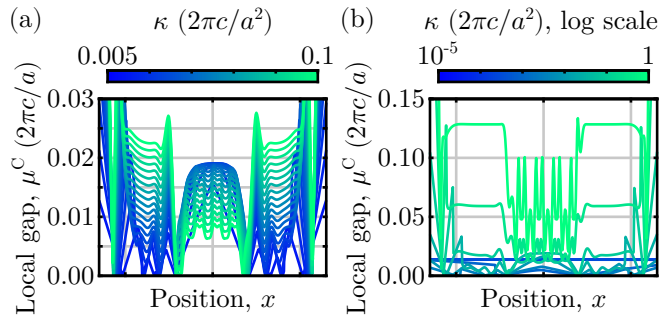}
    \caption{(a) Local gap $\mu^{\textrm{C}}$ calculated for the system from Fig.\ 2 in the main text along the same positions as is shown in Fig.\ 2e (i.e., the green line in Fig.\ 2d). Here, we are using $\omega = 0.37 (2\pi c/a)$. The different colored lines are showing linearly spaced values of $\kappa$. (b) Similar to (a), except for logarithmicly spaced choices of $\kappa = [1 \cdot 10^{-5}, 3.6 \cdot 10^{-5}, 1.3\cdot 10^{-4}, 4.6\cdot 10^{-4}, 0.0017, 0.0060, 0.022, 0.077, 0.28, 1] (2\pi c/a^2)$.}
    \label{fig:kappa}
\end{figure}

However, even though these two limits are boring, for choices of $\kappa$ in between these two limits the spectrum of $L$ can be highly non-trivial, i.e., when the spectral localizer sees information from both the system's Hamiltonian and position operators with relatively equal strength. Moreover, this range of $\kappa$ where $L$ is useful is relatively broad. In Supp.\ Fig.\ \ref{fig:kappa}a, we show the local gap $\mu^{\textrm{C}}$ over the same range of positions as is shown in Fig.\ 2e in the main text, except calculated for values of $\kappa$ that span a factor of $20$. As can be seen, while different values of $\kappa$ yield modestly different values for $\mu^{\textrm{C}}$, the variation in the values of the local gap is significantly reduced to being just a factor of $2$. Moreover, in Supp.\ Fig.\ \ref{fig:kappa}b we show values of $\kappa$ chosen on a logarithmic scale, and we can confirm that the system's topological index is preserved against different choices of $\kappa$ over a range of at least $\kappa = 0.0005 (2\pi c/a^2)$ to $\kappa = 0.077 (2\pi c/a^2)$, i.e.\ a range in excess of two orders of magnitude.

But, if small values of the local gap $\mu^{\textrm{C}}$ are related to the appearance of localized states in the system, would not the changes in the local gap due to different choices of $\kappa$ render this theory useless? No. The reason is that bounds on the approximations discussed after Eq.\ (1) in the main text \emph{also} depend on $\kappa$. In particular, the quantities $[H, \kappa X]$ and $[H, \kappa Y]$ feature prominently in predicting the localization of states when $\mu^{\textrm{C}} = 0$ \cite[Prop.\ II.4]{cerjan_quadratic_2022}. Thus, while different values of $\kappa$ change $\mu^{\textrm{C}}$, they also change the bounds in a similar way, yielding a consistent picture.

\section{An efficient method for calculating the local index in non-Hermitian systems}

Numerically, it is important to approach the determination of the local topological index $C_{(x,y,\omega)}^{\textrm{L}}$ with some care, especially in photonic systems. For example, the matrices $X,Y,H$ in the system shown in Fig.\ 2a each have size $\sim 1.2 \cdot 10^{6} \times 1.2 \cdot 10^{6}$ in our discretization. Attempting to calculate all of the eigenvalues of $L_{(x,y,\omega)}(X,Y,H)$ will cause most computers to run out of memory.

In Hermitian systems, one can make use of Sylvester's law of inertia along with a standard $L_{(x,y,\omega)}(X,Y,H) = N D N^\dagger$ decomposition (called the LDLT decomposition) to substantially speed up this process \cite{sylvester_xix_1852,higham_sylvesters_2014}. In particular, $\textrm{sig}(L_{(x,y,\omega)}) = \textrm{sig}(D)$, i.e., the signature of the spectral localizer is preserved under this form of factorization. Thus, as there are fast sparse methods available to perform LDLT decompositions, it is relatively easy to find a system's local index.

However, this presents a challenge in non-Hermitian systems where Sylvester's law of inertia no longer applies. Instead, to avoid needing to calculate the full spectrum of $L$, we start by turning off the absorption from the stretched-coordinate perfectly matched layer (PML), yielding a Hermitian system where the previous method works. In particular, as our PML is implemented as
\begin{equation}
    \frac{\partial}{\partial x} \rightarrow \frac{1}{1 + \frac{i \sigma_{\textrm{max}}}{\omega}\left(\frac{x}{L_{\textrm{PML}}} \right)^3} \frac{\partial}{\partial x}
\end{equation}
inside the absorbing boundary, where $\sigma_{\textrm{max}}$ is the maximum absorption achieved and $L_{\textrm{PML}}$ is the length of the absorbing layer, the system can be made Hermitian by setting $\sigma_{\textrm{max}} = 0$ (yielding a system bounded by Dirichlet boundary conditions). A similar formula is used for the PML in $y$.

\begin{figure}[h]
    \centering
    \includegraphics{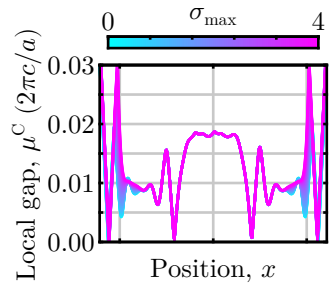}
    \caption{Local gap $\mu^{\textrm{C}}$ calculated for the system from Fig.\ 2 in the main text along the same positions as is shown in Fig.\ 2e (i.e., the green line in Fig.\ 2d). Here, we are using $\omega = 0.37 (2\pi c/a)$ and $\kappa = 0.04 (2 \pi c /a^2)$. The different colored lines are showing linearly spaced values of $\sigma_{\textrm{max}}$ between $0$ and $4$, where $\sigma_{\textrm{max}} = 4$ is the value used in all of the other simulations of the photonic system in this work.}
    \label{fig:sigma}
\end{figure}

Thus, we can determine a non-Hermitian system's local index by starting with a related Hermitian system whose topological markers can be efficiently calculated, slowly turning on the non-Hermiticity, and monitoring the local gap $\mu^{\textrm{C}}$ to ensure it remains open. Of course, if $\mu_{(x,y,\omega)}^{\textrm{C}} \ne 0$ along this path of different systems, the index at $x,y,\omega$ cannot change as no eigenvalues could have crossed the imaginary axis. For the system we consider in the main text in Fig.\ 2, we show this evolution of the local gap as a function of the boundary's absorption in Supp.\ Fig.\ \ref{fig:sigma}. As can be seen, the introduction of the absorption does not yield any new locations where $\mu^{\textrm{C}} = 0$, meaning that the index of the topological photonic crystal in the center of the system remains the same as its index in a Hermitian version of the system.

\section{Additional details on the crystal disorder}

\begin{figure}[h]
    \centering
    \includegraphics{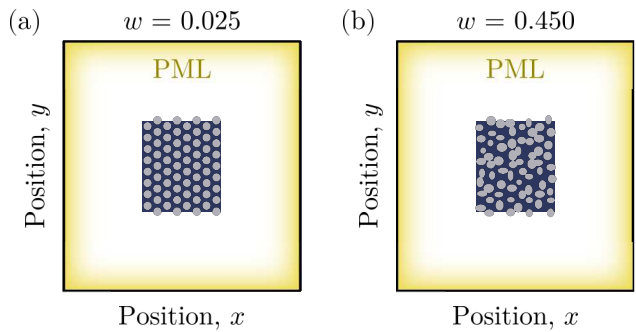}
    \caption{Schematics showing the geometry of the disordered systems discussed in Fig.\ 3 of the main text, with geometric disorder strength $w = 0.025$ and $w = 0.45$. These schematics are shown on the same scale as Fig.\ 2a of the main text.}
    \label{fig:disorderDistribution}
\end{figure}

In Fig.\ 3 of the main text, the local gap $\mu^{\textrm{C}}$ and local invariant $C^{\textrm{L}}$ are shown for two disorder configurations with different disorder strengths. In Supp.\ Fig.\ \ref{fig:disorderDistribution}, we show the dielectric distribution for these two disorder configurations. These configurations are generated by changing the width, length (major and minor ellipse axes), and position of the dielectric rods (which are circular in the ordered system). In particular, for each rod we generate four uniformly distributed random numbers $\xi_j \in [-0.5, 0.5]$. Two of these random numbers are used to shift the $x$ and $y$ coordinates of the rod's center, $x_{\textrm{new}} = x_0 + w \xi_1$ and $y_{\textrm{new}} = y_0 + w \xi_2$, respectively. The other two random numbers treat the rod as being an ellipse, and change the length of its major and minor axes by $w \xi_j/2$. Here, $w$ parameterizes the geometric strength of the disorder. However, given that the inverse square root of $\overline{\overline{\varepsilon}}$ is what appears in the system's Hamiltonian, see Eq.\ (4) in the main text, there is no simple relationship between $w$ and $\Vert \Psi^\dagger \delta H \Psi \Vert$.

\begin{figure}[h]
    \centering
    \includegraphics{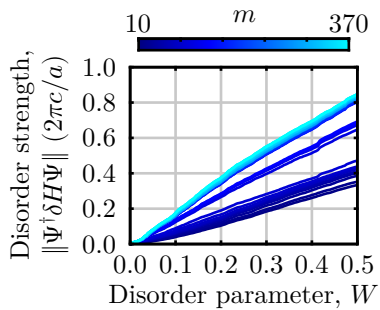}
    \caption{Dependence of the disorder strength $\Vert \Psi^\dagger \delta H \Psi \Vert$ on $w$, the magnitude of the random numbers chosen to implement the disorder for different choices of $m$, the number of retained eigenvectors in $\Psi$. The choices of $m$ are linearly spaced between $10$ and $370$ in steps of $10$.}
    \label{fig:disStr}
\end{figure}

Additionally, in the calculation of the strength of these disorder configurations, one must make a choice for how many eigenvectors of the ordered system $H$ to retain in $\Psi$ for calculating $\Vert \Psi^\dagger \delta H \Psi \Vert$ (this number of retained eigenvectors is $m$ in the relevant discussion surrounding Fig.\ 3 in the main text, while the total number of possible eigenvectors is $n$). Of course, $\Vert \Psi^\dagger \delta H \Psi \Vert$ must be dependent on $m$ --- clearly as $m \rightarrow n$, $\Vert \Psi^\dagger \delta H \Psi \Vert \rightarrow \Vert \delta H \Vert$ as the $\ell^2$ matrix norm is basis independent. However, for a broad range of choices of $m \ge 180$ that retain only a few hundred eigenvectors of $H$, we find that $\Vert \Psi^\dagger \delta H \Psi \Vert$ is nearly independent of $m$, see Supp.\ Fig.\ \ref{fig:disStr}. This justifies our choice of spectral truncation for the disorder strength calculation, and the values quoted in the main text use the largest $m$ shown in Supp.\ Fig.\ \ref{fig:disStr}, $m = 370$. For this value of $m$, a disorder parameter of $w = 0.025$ corresponds to a disorder strength of $\Vert \Psi^\dagger \delta H \Psi \Vert = 0.026 (2\pi c/a)$, and $w = 0.45$ corresponds to $\Vert \Psi^\dagger \delta H \Psi \Vert = 0.77 (2\pi c/a)$. In the main text, these values are compared against the local gap at the center of the ordered system, $\mu_0^{\textrm{C}} = 0.0185 (2\pi c/a)$, which is calculated at $\omega = 0.37 (2\pi c/a)$ using $\kappa = 0.04 (2\pi c/a^2)$.

\section{Classifying the topology of gapless tight-binding heterostructures}

While the main text focuses on applications of the spectral localizer to non-Hermitian photonic systems the spectral localizer can be applied to any crystalline system. To demonstrate the broader applications of these methods we consider examples of gapped and gapless topological heterostructures in tight-binding models. Note, unlike the photonic system discussed in the main text, all of tight-binding models we discuss here are Hermitian with open boundary conditions. 

\begin{figure*}
    \centering
    \includegraphics{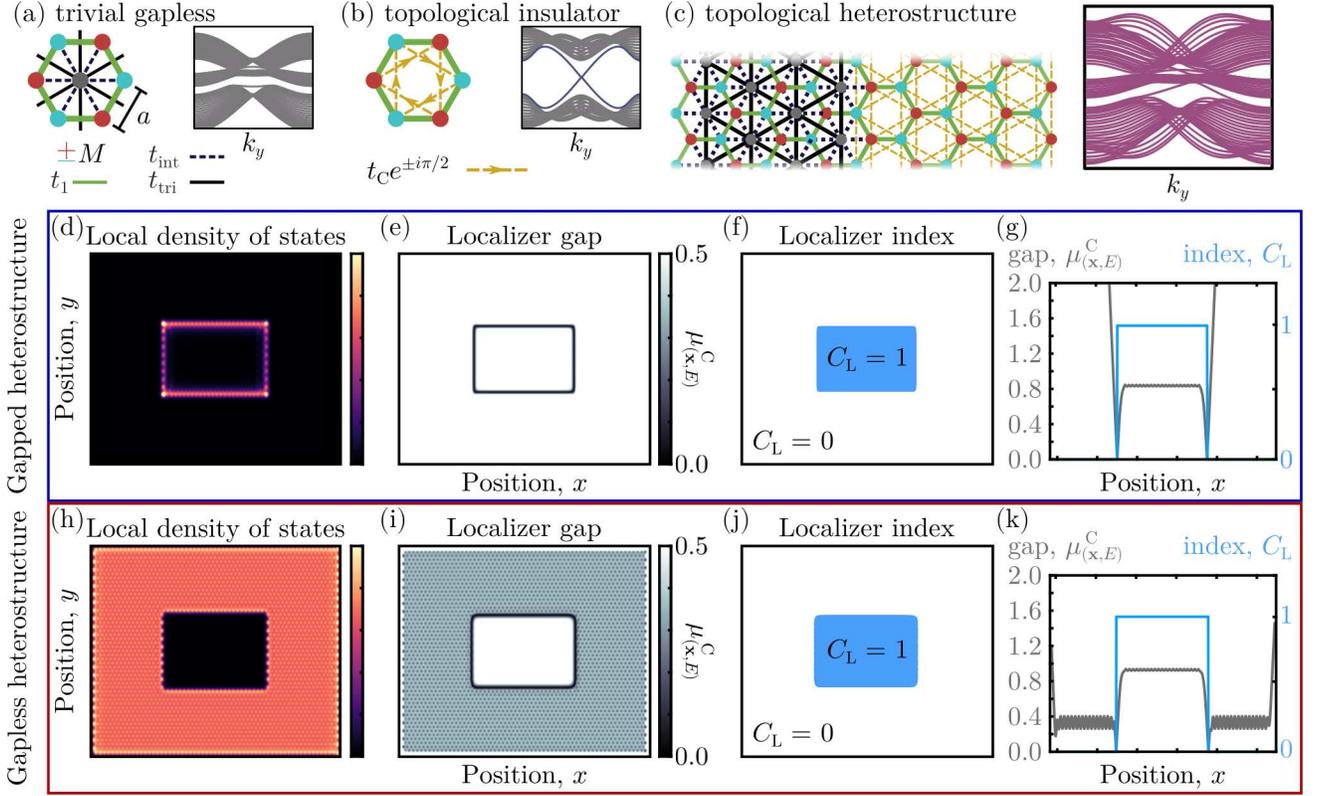}
    \caption{(a) Schematic and ribbon band structure for a honeycomb lattice (red and blue vertices) with an interstitial triangular lattice (gray vertices), with honeycomb site couplings $t_1$, triangular site couplings $t_{\textrm{tri}}$, and honeycomb-triangle couplings $t_{\textrm{int}}$. The ribbon band structure is calculated using $t_1 = t$, $t_\textrm{C} = 0$, $t_{\textrm{tri}} = 0.2t$, $t_{\textrm{int}} = 0.3t$, $M = \pm \sqrt{3}/2$, and $M_{\textrm{tri}} = 0$. (b) Schematic and ribbon band structure of a Haldane topological insulator with direction-dependent next-nearest neighbor couplings $t_{\textrm{C}}e^{\pm i \pi/2}$. The ribbon band structure is calculated using $t_1 = t$, $M = 0$, $t_{\textrm{C}} = 0.5t$, $\phi = \pi/2$. (c) Schematic and ribbon band structure for a heterostructure formed between the lattices in (a) and (b).
    (d-f) LDOS (d), localizer gap  (e), and local index (f) at $E = 0$ for a heterostructure formed by a trivial insulating Haldane lattice described by Eq.\ (\ref{eq:Hhal}) with $t_1 = t$, $t_\textrm{C} = 0$, and $M = \pm 4\sqrt{3}/2$, surrounding a Haldane topological insulating lattice with $t_1 = t$, $M = 0$, $t_{\textrm{C}} = 0.5t$, $\phi = \pi/2$. All three plots are shown on the same spatial scale, with a system surrounded by open boundary conditions. (g) Horizontal line cut of the localizer gap and index through the system's center at $E = 0$. (h-k) Similar to (d-g), except with a gapless outer lattice described by Eq.\ (\ref{eq:HhalMetal}) with $t_1 = t$, $t_\textrm{C} = 0$, $t_{\textrm{tri}} = 0.2t$, $t_{\textrm{int}} = 0.3t$, $M = \pm \sqrt{3}/2$, and $M_{\textrm{tri}} = 0$. For all calculations using the spectral localizer, $\kappa = t/a$ is used.}
    \label{fig:2}
\end{figure*}

To begin, we must first assemble the topological heterostructures. A standard choice of topological insulator is the Chern insulator realized via Haldane's model~\cite{haldane_model_1988}. The Haldane lattice has the following Hamiltonian:
\begin{align}
    H ~=~& M \sum_{m,n} \left(a_{m,n}^\dagger a_{m,n} - b_{m,n}^\dagger b_{m,n} \right) \notag\\
    &- t_1 \sum_{\langle (m,n),(m',n')\rangle} \left(b_{m',n'}^\dagger a_{m,n} + a_{m,n}^\dagger b_{m',n'} \right) \notag \\
    & -t_\textrm{C} \sum_{\langle \langle (m,n),(m',n')\rangle \rangle} \left(e^{i \phi} a_{m',n'}^\dagger a_{m,n} + e^{-i \phi} a_{m,n}^\dagger a_{m',n'} + e^{i \phi} b_{m',n'}^\dagger b_{m,n} + e^{-i \phi} b_{m,n}^\dagger b_{m',n'} \right). \label{eq:Hhal}
\end{align}
This model has two sites per unit cell that form a hexagonal lattice, with annihilation (creation) operators of these sites given in the $(m,n)$th unit cell by $a_{m,n}$ and $b_{m,n}$ ($a_{m,n}^\dagger$ and $b_{m,n}^\dagger$) with nearest neighbor coupling $t_1$, onsite energy $M =0$, and next-nearest neighbor coupling $t_C$. Here, $\langle \rangle$ denotes a sum over nearest neighbors, while $\langle \langle \rangle \rangle$ denotes the sum is over next-nearest neighbors. To push the Haldane lattice into one of its topological phases, we set $t_1 = t$, $M = 0$, $t_C = t/2$, and $\phi = \pi/2$, see Fig.\ \ref{fig:2}b. For a strip of this material we can calculate the ribbon band structure as seen in Fig.\ \ref{fig:2}b, where the chiral edge states are clearly visible crossing through the system's bulk band gap. However, to form topological heterostructures, we need to create a material to interface with this topological insulator. 

To simplify the choice of interface coupling we use a Haldane lattice in its trivial phase as the topologically distinct insulator and a trivialized metallized Haldane lattice for the topologically distinct gapless material. To trivialize either Haldane system we set $t_C = 0$ and set the onsite energy $M= \pm 2 \sqrt{3}$ for the trivial insulator and $M = \pm \sqrt{3}/2$ for the trivial metal.
To metallize the Haldane system we couple it to a trivial metallic triangular lattice (so that there are now 3 sites per unit cell, i.e., the grey, red, and blue sites in Fig.\ \ref{fig:2}), with an inter-lattice coupling strength $t_{\textrm{int}}= 0.3t$ \cite{bergman_bulk_2010,junck_transport_2013}. We set the coupling strength within the triangular lattice to be $t_{\textrm{tri}} = 0.2 t$. Altogether, the Hamiltonian for a metallized Haldane lattice is written as
\begin{align}
    H ~=~& M \sum_{m,n} \left(a_{m,n}^\dagger a_{m,n} - b_{m,n}^\dagger b_{m,n} \right) + M_{\textrm{tri}} \sum_{m,n} \left(c_{m,n}^\dagger c_{m,n} \right) \notag \\
    &- t_1 \sum_{\langle (m,n),(m',n')\rangle} \left(b_{m',n'}^\dagger a_{m,n} + a_{m,n}^\dagger b_{m',n'} \right) -  t_{\textrm{tri}} \sum_{\langle (m,n),(m',n')\rangle} \left(c_{m',n'}^\dagger c_{m,n} + c_{m,n}^\dagger c_{m',n'} \right) \notag \\
     &-t_\textrm{C} \sum_{\langle \langle (m,n),(m',n')\rangle \rangle} \left(e^{i \phi} a_{m',n'}^\dagger a_{m,n} + e^{-i \phi} a_{m,n}^\dagger a_{m',n'} + e^{i \phi} b_{m',n'}^\dagger b_{m,n} + e^{-i \phi} b_{m,n}^\dagger b_{m',n'} \right) \notag \\
     &- t_{\textrm{int}} \sum_{\langle (m,n),(m',n')\rangle} \left(a_{m',n'}^\dagger c_{m,n} + c_{m,n}^\dagger a_{m',n'} + b_{m',n'}^\dagger c_{m,n} + c_{m,n}^\dagger b_{m',n'} \right). \label{eq:HhalMetal}
\end{align}
Here, $c_{m,n}$ ($c_{m,n}^\dagger$) is the annihilation (creation) operator for the triangular lattice site in the $(m,n)$th unit cell. A strip of the trivial metallic Haldane lattice is used to calculate the ribbon band structure in Fig.\ \ref{fig:2}a. 

Concatenating both systems from Fig.\ \ref{fig:2}a and Fig.\ \ref{fig:2}b into a gapless topological heterostructure forms a system with a band structure that reproduces the challenge associated with gapless topological heterostructures discussed in the main text, see Fig.\ \ref{fig:2}c. In the band structure calculations, the bands corresponding to interface-localized states are obscured by the degenerate bulk bands from the interstitial triangular lattice. Within this energy range, it is not possible to use topological band theory to identify the existence of interface-localized chiral modes --- the heterostructure does not exhibit a common bulk band gap between the two constituent materials so band theory cannot be used to predict a measure of protection, and the edge states cannot be uniquely identified at some energy and wavevector. 

The challenges involved in identifying the topology of the gapless heterostructure remain apparent in the system's local density of states (LDOS). In Supp.\ Fig.\ \ref{fig:2}d and h, we show the LDOS within the topological insulator's bulk band gap $E=0$ for finite gapped and gapless heterostructures, respectively. For both systems, the inner material is a topological insulator, the outer material is topologically trivial, and the outer-most boundary of the outer material has open (Dirichlet) boundary conditions. While the chiral edge states at this energy can be clearly identified in the gapped heterostructure's LDOS, Supp.\ Fig.\ \ref{fig:2}d, the states due to the interstitial triangular lattice's band in Supp.\ Fig.\ \ref{fig:2}h prohibit the numerical observation of any chiral edge states. 

We can use the spectral localizer \cite{loring_k-theory_2015,loring_finite_2017,loring_spectral_2020} to prove that the gapless topological heterostructure in Fig.\ \ref{fig:2}h-k still possesses boundary-localized resonances that are connected to the non-trivial topology of the central lattice. For a $2$-dimensional Hermitian tight-binding model, the local marker and protection at position $\mathbf{x}$ and energy $E$ are found by first forming the spectral localizer:
\begin{equation}
{L}_{(\mathbf{x},E)} (\mathbf{X},{H}) =
 \kappa ({X}_1-x_1 {I}) \otimes {\sigma}_x +\kappa ({X}_2-x_2 {I}) \otimes {\sigma}_y + ({H} - E {I}) \otimes {\sigma}_{z}, \label{eq:L}
\end{equation}
where $\mathbf{X} = (X_1,X_2)^T$, ${X}_j$ is the $j^{th}$ position operator,$\mathbf{x} = (x_1,x_2)^T$, ${x}_j$ is the $j^{th}$ position coordinate, $\sigma_j$ are the Pauli spin matrices, $\kappa$ is a positive scaling coefficient with units of energy times inverse distance, and $I$ is the identity matrix.

Using the spectral localizer gap and index, we resolved the local Chern numbers for both topological heterostructures shown in Fig.\ \ref{fig:2}, and numerically observe that all substructures are comprised of materials with different invariants in their bulk. Moreover, for the gapless system, applying the spectral localizer and plotting the local gap versus position shows the closing of $\mu_{(\mathbf{x},E)}^\textrm{C}$ near the heterostructure's interface where the local marker changes, which requires the system to possess topological edge resonances even if they are obscured in the LDOS. Finally, we note that the small, but non-zero, local gap $\mu_{(\mathbf{x},E)}^\textrm{C}$ on both sides of the heterostructure's interface guarantees the edge resonance's protection against modest system perturbations. As tight binding models are not unbounded operators, they are not subject to the considerations discussed in the main text surrounding the difficulty of defining the strength of a perturbation through its matrix norm. Instead, we are left with the standard topological protection predicted by the spectral localizer: a perturbation $\delta H$ that is weaker than the local gap, $\Vert \delta H \Vert < \mu_{(\mathbf{x},E)}^\textrm{C}$, cannot change the system's local topology at $(\mathbf{x},E)$ \cite{loring_k-theory_2015}.

\begin{figure}
    \centering
    \includegraphics{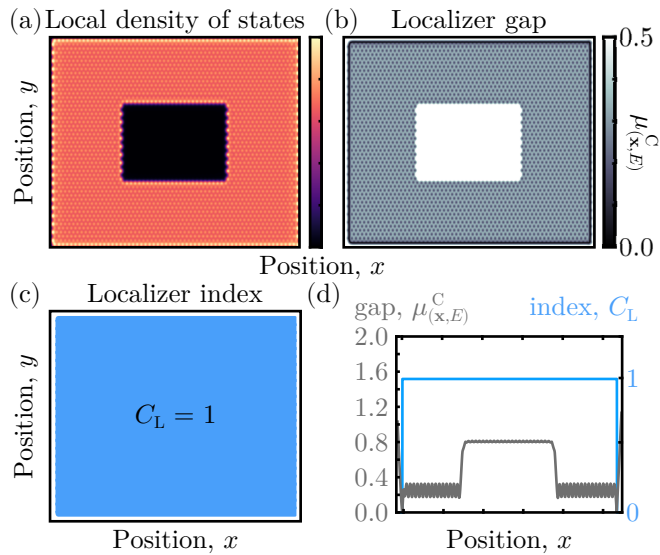}
    \caption{(a-c) Show the LDOS (a), localizer gap (b), and localizer index (c) at $E=0$ for a gapless heterostructure formed with an inner Haldane topological insulator (same as Fig.\ \ref{fig:2}), and a gapless topological outer lattice described by Eq.\ (\ref{eq:HhalMetal}) that incorporates direction-dependent next-nearest neighbor couplings with $t_1 = t$, $t_{\textrm{tri}} = 0.2t$, $t_{\textrm{int}} = 0.3t$, $M = 0$, $M_{\textrm{tri}} = 0$, $t_{\textrm{C}} = t/2$, and $\phi = \pi/2$. All three plots are shown on the same spatial scale, with open boundary conditions. (d) Horizontal line cut of the localizer gap and index through the system's center and at $E = 0$. For all calculations using the spectral localizer, $\kappa = t/a$ is used.}
    \label{fig:3}
\end{figure}

To further prove that the interface-localized resonance in a gapless heterostructure can be attributed to the change in the materials' bulk invariants, i.e., that bulk-boundary correspondence still holds for gapless heterostructures, we slightly modify the gapless material on the exterior of the heterostructure to possess the same bulk invariant as the central insulating material, see Fig.\ \ref{fig:3}. This is constructed from the same topological Haldane lattice that is then coupled to the trivial gapless triangular lattice. The result is a gapless topological structure with degenerate bulk and a topological edge state protected by a local gap \cite{cerjan_top_metal_2022}.

Now, we see that the system is no longer a topological heterostructure; the local gap $\mu_{(\mathbf{x},E)}^\textrm{C}$ no longer closes at the interface between the two materials, indicating the disappearance of the topological interface-localized resonances, while the local maker is seen to be uniform across the entire crystal. Furthermore, the local gap closes around the system's perimeter due to the use of open boundary conditions, indicating a topological phase transition between the outer lattice material and the surrounding (insulating) vacuum.

\bibliography{metal_insulator_interface}